\documentclass{aa}
\def\et{et al.}                        

\begin{document}

\title{Strongly decelerated expansion of SN\,1979C}


\author{J.M.\ Marcaide\inst{1}
  \and
   M.A.\ P\'{e}rez-Torres\inst{1,2}
  \and
   E.\ Ros\inst{3}
  \and
   A.\ Alberdi\inst{4}
  \and
   P.J.\ Diamond\inst{5}
  \and
   J.C.\ Guirado\inst{1}
  \and
   L. Lara\inst{4}
  \and
   S.D.\ Van Dyk\inst{6}
  \and
   K.W.\ Weiler\inst{7}
  }

\offprints{J.M. Marcaide,
      email: J.M.Marcaide@uv.es}

\institute{Departamento de Astronom\'{\i}a, Universitat de
      Val\`encia, E-46100 Burjassot, Spain 
 \and
     Istituto di Radioastronomia/CNR, Via P.\ Gobetti 101,
     I-40129 Bologna, Italy 
 \and
     Max-Planck-Institut f\"ur Radioastronomie,
     Auf dem H\"ugel 69, D-53121 Bonn, Germany 
 \and
     Instituto de Astrof\'{\i}sica de Andaluc\'{\i}a, CSIC,
     Apdo.\ Correos 3004, E-18080 Granada, Spain 
 \and
     MERLIN/VLBI National Facility, Jodrell Bank Observatory,
     Macclesfield, Cheshire SK11 9DL, UK
 \and
     Infrared Processing and Analysis Center, California
     Institute of Technology, Mail Code 100-22, Pasadena,
     CA 91125, US 
 \and
     Remote Sensing Division, Naval Research Laboratory,
     Code 7213, Washington, DC 20375-5320, US 
     }
\date{}

\abstract{ We observed \object{SN\,1979C} in \object{M100} on 4
June 1999, about twenty years after explosion, with a very
sensitive four-antenna VLBI array at the wavelength of
$\lambda$18\,cm. The distance to M100 and the expansion velocities
are such that the supernova cannot be fully resolved by our
Earth-wide array. Model-dependent sizes for the source have been
determined and compared with previous results.  We conclude that
the supernova shock was initially in free expansion for
6$\pm2$\,yrs and then experienced a very strong deceleration. The
onset of deceleration took place a few years before the abrupt
trend change in the integrated radio flux density curves. We
estimate the shocked swept-up mass to be $M_{\rm sw} \sim1.6
M_{\sun}$, assuming a standard density profile for the CSM. Such a
swept-up mass for \object{SN\,1979C} suggests a mass of the
hydrogen-rich envelope ejected at explosion no larger than $M_{\rm
env} \sim 0.9 M_{\sun}$. If  \object{SN\,1979C} originated in a
binary star the low value of $M_{\rm env}$ suggests that the
companion of the progenitor star stripped off most of the
hydrogen-rich envelope mass of the presupernova star prior to the
explosion. 
}

\maketitle

\section{Introduction \label{introduction}}

Supernova \object{SN\,1979C} in \object{M100} was discovered on
1979 April 19 by Gus E.\ Johnson (Mattei \et\ \cite{mat79}) at
around magnitude 12 in the visible, and it is thought to have
exploded on April 4, 1979 (Weiler \et\ \cite{wei86}). Optical
spectra first showed a featureless continuum, which later evolved
to exhibit strong H$\alpha$ emission (e.g., Schlegel
\cite{sch96a}). \object{SN\,1979C} was extraordinarily luminous,
$M_B^{\rm max}\approx -20$ (e.g., Young \& Branch \cite{you89}),
making it among the most luminous type II supernova ever observed.
\object{SN\,1979C} was also found to be a radio supernova
(RSN)(Weiler and Sramek, 1980). Indeed, at a distance of
$16.1\pm1.3$\,Mpc estimated for \object{M100} (Ferrarese \et\
\cite{fer96}) it is one of the intrinsically strongest RSN (Weiler
\et\ \cite{wei96}). From H$\alpha$ widths an expansion speed of
9\,200$\pm$500\,km\,s$^{-1}$ was estimated (Panagia \et\
\cite{pan80} and G. Vettolani, priv. comm.) for epoch around
45\,d.

\object{SN\,1979C} has been classified as a SN of type II-L. Fesen
\et\ (1999), hereafter F99, have studied the late-time optical
properties of \object{SN\,1979C} and other three type II-L
supernovae. Late-time optical emission lines from type SNe II-L
appear surprisingly steady over long time intervals. These authors
find similar expansion velocities in the range
5100--6000\,km\,s$^{-1}$ in the strongest lines present
(H$\alpha$, [{\sc Oi}] 6300,~6364\,{\AA}, [{\sc Oii}] 7319,~7330\,{\AA},
and [{\sc Oiii}] 4959,~5007\,{\AA}) for all type II-L supernovae. From
the asymmetries of the lines these authors find evidence of dust
formation. Strong radio emission from a SN appears to be strongly
correlated with late time bright optical emission, with SN\,1979C
as a prime example. In \object{SN\,1979C}, all detected lines but
H$\alpha$ show emission peaks at $-5\,000$ and
$-1\,000$\,km\,s$^{-1}$ suggestive of clumpy emission from
material coming mostly from ejecta, and physically separated from
the material emitting H$\alpha$. The H$\alpha$ emission tracks
well the peak 6\,cm emission (Weiler \et\ \cite{wei91}) indicating
that both emissions are related to the material from the swept-up
shell.

The evolution of the radio emission from SN\,1979C has been
interpreted within the mini-shell model (Chevalier \cite{che82})
as due to synchrotron emission from the outer part of the shock
swept-up shell of circumstellar material attenuated by thermal
absorption from the even more distant ionized CSM (Weiler \et\
\cite{wei86}, Montes \et\ \cite{mon00}).  Synchrotron self
absorption which has been shown to be relevant in
\object{SN\,1993J} (Fransson \& Bj\"ornsson \cite{fra98},
P\'{e}rez-Torres \et\ \cite{per01}) might also play a role in
\object{SN\,1979C}, although Chevalier (\cite{che98}) is not of
this opinion. The modulation present in the radio light curves of
\object{SN\,1979C} has led Weiler \et\ (\cite{wei92}) and Montes
\et\ (\cite{mon00}) to model the progenitor as possibly being in a
detached eccentric binary system with a less massive companion.
>From a study of the emission environment of \object{SN\,1979C},
Van Dyk \et\ (\cite{vdy99}) have estimated the mass of the
progenitor to be $17-18\,(\pm3)$\,M$_\odot$. Hydro-dynamical
simulations by Schwarz \& Pringle (\cite{sch96}) confirm that a
binary system similar to that proposed by Weiler \et\
(\cite{wei92}) with a spiral shaped stellar wind around the
progenitor is feasible for SN\,1979C. Such modulation of the
density of the gas surrounding the RSN and now being shocked,
might have dramatic effects on the shape of the radio structure. A
determination of such radio structure would be a powerful way to
further understand the pre-supernova phase of the progenitor of
\object{SN\,1979C}.

\begin{figure}
\vspace{180pt} \includegraphics{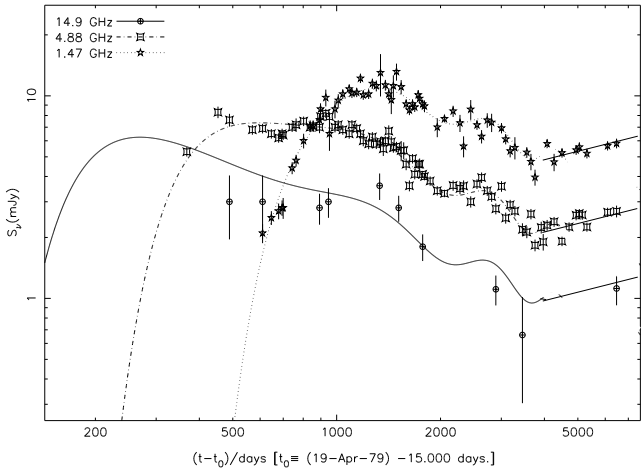} \caption{Radio light curves for SN\,1979C in
M100 at wavelengths of 20, 6, and 2\,cm. The curves represent the
best-fit model light curves. The best-fit parameters were
determined using only data through 1990 December (day $\sim$4300).
Notice the flux density increase from day 4000 onwards (Montes et
al. 2000). \label{fig:lightcurve}}
\end{figure}

The radio emission of this RSN, monitored by Montes \et\
(\cite{mon00}), has started to show an interesting trend which can
be seen in Fig.\ \ref{fig:lightcurve}: after years of steady
decline the radio emission has been constant, or perhaps
increasing, since about 1990 ($\sim10$\,yr) in a way somewhat
reminiscent of the radio emission increase for the case of
\object{SN\,1987A} (Ball \et\ \cite{ball95}). It appears that
after a standard RSN radio emission decline phase (Weiler \et\
\cite{wei96}) the expanding shock may have encountered denser
ionized gas which has boosted the radio emission. These results
may be in accord with those of \cite{fes99} which show that high
electron densities (10$^{5-7}$ cm$^{-3}$) are necessary to explain
the absence of several lines, such as [{\sc Oii}]
3726,~3729\,{\AA} in their optical spectra.

Soft X-rays have been detected from \object{SN\,1979C} by Immler
\et\ (\cite{imm98}), Ray \et\ (\cite{ray01}), and Kaaret
(\cite{kaa01}). The soft X-ray luminosity appears constant or
slightly declining from epoch 16\,yr through 20\,yr. This result
appears compatible with the model predictions of circumstellar
interactions (Chevalier \& Fransson \cite{che94}).

VLBI observations at 3.6, 6, 13 and 18\,cm by Bartel \et\ (1985),
hereafter B85, determined a size of \object{SN\,1979C} and its
growth for epochs ranging from Dec~1982 to May~1984 assuming a
synchrotron emitting optically thin spherical model. At those
early epochs the source could not be resolved by any VLBI array.
At an average expansion speed of, for example, $\sim$8\,000
km\,s$^{-1}$, one should expect a shell radius growth rate of
$\sim$ 0.3$\mu$as\,day$^{-1}$ which over 20 years should yield a
shell radius of $\sim$2.2\,mas which could be resolved by
$\lambda$6\,cm VLBI if detected on the long baselines. However,
due to sensitivity limitations of the available arrays, the
chances of detecting it at $\lambda$\,6\,cm at the present time
are null. On the other hand, at $\lambda$18\,cm the emission
strength of \object{SN\,1979C} and the sensitivity of the arrays
make the detection possible. We have carried out state-of-the-art
$\lambda$18\,cm observations in an attempt to learn as much as
possible about the source structure.

\section{Observations \label{sect:observations}}

\object{SN\,1979C} was observed on 4 June 1999 from 19:30\,UT to
01:30\,UT at 1.6\,GHz with a sensitive transatlantic VLBI array.
The antennas used were (diameter, location): Effelsberg (100m,
Germany), Robledo (70m, Spain), phased-VLA (130m-equivalent, NM,
US), and Goldstone (70m, CA, US).  Data were recorded using a
standard VLBA mode with a data rate of 256Mb/s, covering a
bandwidth of 64\,MHz. The data were correlated at the VLBA
Correlator Center of the National Radio Astronomy Observatory
(NRAO)\footnote{NRAO is operated under cooperative agreement with
the National Science Foundation.}in Socorro, NM, US.

The observations were made with a  duty cycle of about 26\,min
consisting of 3\,min observations
of the calibrator source \object{PKS\,B1157+156}, followed by 5\,min
observations of the calibrator source \object{TXS\,1214+161}, in turn followed
by 14\,min observations of the source SN\,1979C.  The source
\object{3C\,274} was observed twice
as a calibrator, once at the beginning and once at the end of the
observations.

The correlated data were analyzed using the Astronomical Image
Processing System ({\sc aips}). A basic integration time of 2\,sec
was used. The phase slopes across each of the 8\,MHz bands and
over the 64\,MHz synthesized band were aligned using a phase
calibration determined from observations of the calibrator source
\object{3C\,274}. The data of the calibrators
\object{PKS\,B1157+156} and \object{TXS\,1214+161} were then
fringe-fitted in a standard manner.  In the next step, phases,
delays, and delay-rates corresponding to the sources
\object{PKS\,B1157+156} and \object{TXS\,1214+161} were interpolated
to the source  \object{SN\,1979C}, to provide an initial model for
the fringing of the data of this radio source. A solution was
obtained by fringing the data using narrow search windows set
around the initial model. Fringe detections of SN\,1979C were
obtained for all baselines at almost all times. However, some
SN\,1979C data had to be removed from the analysis after visual
inspection of the data of the two main calibrators. Indeed, for
some time ranges in the observations there were calibration
problems or data with lower than acceptable quality. For instance,
high winds at the phased-VLA produced an auto-stowing of some
antennas and problems in the phasing during the second half of the
observations. Therefore, data from the VLA from 23:30 UT to 01:15
UT were edited.

Adjacent to the VLBI observations reported in this paper, some cross-scans on
\object{PKS\,B1157+156} were carried out at 1.6\,GHz to fix the
pointing parameters of the antenna in Effelsberg. As a byproduct,
those observations allowed the total flux density of
PKS\,B1157+156 to be measured as 0.79$\pm$0.02\,Jy (A.\,Kraus,
priv.comm.). This value was adopted to set our flux density scale.
Then, by readjusting the gains of the antennas, the VLBI map of
PKS\,B1157+156 was forced to have the same flux density as the
total flux density (that is, PKS\,B1157+156 was assumed unresolved
by our VLBI array.) With this calibration procedure for the
antennas of the array a total flux density of 5.4$\pm$0.2\,mJy was
obtained for \object{SN\,1979C}.

Further model fitting analysis was carried out with {\bf {\sc
modelfit}} in {\sc difmap} (Shepherd \et\ \cite{she94}), and with
{\bf {\sc qfit}} in the Caltech VLBI Analysis Package (Pearson \&
Readhead \cite{pea84}).

\section{The size of \object{SN\,1979C} \label{sect:size79c}}

We assumed spherical symmetry in the modeling.  The data remaining
after the editing were of the highest quality, but from a look at
the data it was not fully clear that all points in the visibility
vs.\ uv-radius plot are on the main lobe of the interferometric
response.  We had to prove to ourselves that such was indeed the
case.  Good measurements of the closure phases (all being around
zero and none around 180$^\circ$) were a great help. We conducted
several tests and model fits to convince ourselves that indeed the
array at $\lambda$18\,cm was not large enough to be able to
resolve the structure of SN\,1979C. Unfortunately, there are then
many models which can reproduce the data equally well (see
Marscher \cite{mar85} and \cite{bar85}).

Thus, we decided to determine the size of SN\,1979C for the
following putative source shapes: an optically thick source, an
optically thin shell of width 30\% of the external radius, and a
ring that is an infinitely bright shell of zero width.  The last
one is obviously unphysical and its determination is only valuable
as a limit. Even the first of the three putative source shapes is
not expected given that the spectrum of the source is optically
thin with $S\propto\nu^{-0.63\pm0.03}$ (Montes \et\ \cite{mon00}).
The diameters determined by model fitting from our data for such
models are $4.57\pm0.25$, $3.60\pm0.17$, and $3.10\pm0.14$\,mas,
respectively. The uncertainties were approximately estimated from
the statistical errors given by the programs {\sc qfit} and {\sc
modelfit}, taking into account discrepant results yielded by them.

\begin{figure}[htbp]
\vspace{174pt} \includegraphics{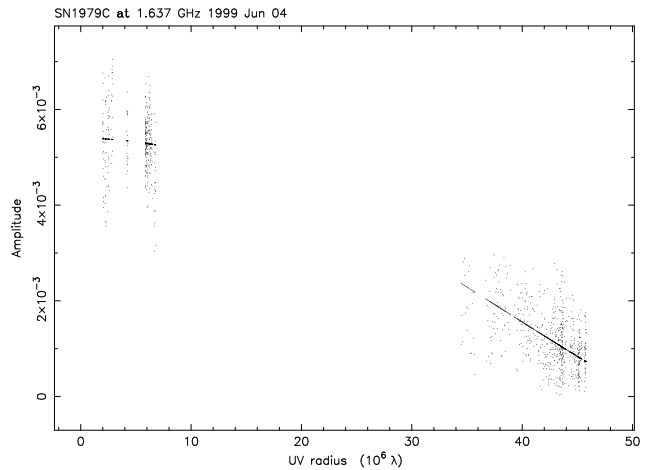} \caption{Visibility amplitude versus
interferometric distance.  The model (solid line) of 3.60\,mas
outer diameter shell with a 30\% thickness is plotted superimposed
on the data. \label{fig:radplot}}
\end{figure}

A fit for the preferred model of an optically thin shell of
diameter $3.60\pm0.17$\,mas  with width 30\% of the external
radius is shown, along with the measured visibilities, as the
solid line in Fig.\ \ref{fig:radplot}.

\section{Discussion \label{sect:discussion}}

Marcaide \et\ (\cite{mar95a,mar95b,mar97}) determined for
\object{SN\,1993J} a shell structure with a shell width 30\% of
the outer radius, somewhat larger than that predicted by Chevalier
\& Fransson (\cite{che94}). Therefore, we have adopted a shell of
width 30\% of the external radius as a model for SN\,1979C in this
paper and all our discussion will be based on such model. For the
present discussion all previous results by \cite{bar85} will be
transformed (normalized) to this model.

At a distance of 16.1\,Mpc to \object{M100} an angular size of
1\,mas corresponds to a linear size of $2.41\cdot10^{17}$\,cm.
Hence, the source outer radius (1.80\,mas) corresponds to
$4.33\cdot10^{17}\,{\rm cm}$. Since 20.12\,yr have elapsed since
the explosion, the corresponding average supernova expansion
velocity is $\sim$6\,800\,km\,s$^{-1}$.  Since early expansion
speeds were estimated at epoch 45\,d from the H$\alpha$ lines as
$\sim$9\,200\,km\,s$^{-1}$ it follows that the expansion has been
strongly decelerated by the action of the circumstellar matter.
How strongly? It is difficult to make a detailed analysis to
answer such a question, since intermediate size estimates are not
available. Nonetheless, we now discuss our VLBI results, in
conjunction with previously published VLBI and optical
spectroscopy data, to show that \object{SN\,1979C} has indeed
suffered a heavy deceleration in its expansion, starting $6\pm2$
\,yr after its explosion.

B85 determined for SN\,1979C a diameter of
$2.16^{+0.44}_{-0.50}$\,mas 5.17\,yr after explosion, from
1.67\,GHz VLBI observations using an optically thin uniform sphere
model. This value translates into a diameter of
$1.56^{+0.31}_{-0.36}$\,mas for our discussion  shell model. This
size corresponds to an average speed for the first 5.17\,yr of
$\sim$11\,500\,km\,s$^{-1}$. On the other hand, the average speed
of the supernova over the first six weeks after explosion as
determined from the widths of H$\alpha$ lines was $\sim9\,200\pm
500$ \,km\,s$^{-1}$ (Panagia \et\ \cite{pan80}, G. Vettolani,
priv. comm.), which is about 20\% lower than the average velocity
estimated for the first 5 years from the \cite{bar85} estimates
normalized to our discussion model. These two results are nicely
consistent. Indeed, in the mini-shell model of Chevalier
\cite{che82}, the radio emission comes from the forward shock,
while the optical, UV, and X-ray emissions come mainly from the
reverse shock, which moves slower than the forward shock. Hence
the velocity ratio determined from the VLBI and optical
observations is roughly what is expected for the radii ratio of
the forward and reverse shocks, in a model of 20-30\% shell width
(Chevalier \& Fransson \cite{che94}, Marcaide \et\ \cite{mar97})
if \object{SN\,1979C} followed a free, self-similar expansion for
the first 5 years after its explosion. A free expansion for the
first 4 years, followed by a decelerated expansion would also be
compatible with the mentioned results. These results should be
understood as evidence of free expansion of \object{SN\,1979C} for
the first 4-5 years and of the presence of a shell-like structure,
and not as evidence of a given shell width, since different model
normalizations of the results of B85 would yield equally
consistent results and a similar conclusion.

\begin{figure}
\vspace{174pt} \includegraphics{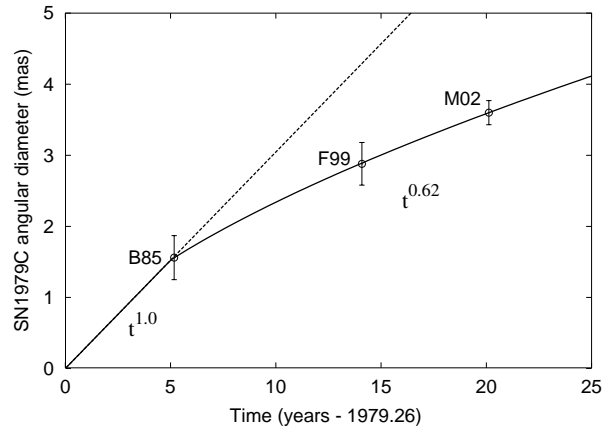} \caption{Angular diameter vs.\ age of
\object{SN\,1979C}. {\bf B85} indicates the size estimate from
Bartel \et\ (1985) normalized to our model (see text), {\bf F99},
the size estimated by us based on the optical results by Fesen
\et\ (1999), and {\bf M02} indicates our results.  The solid line
indicates a possible expansion, which is free for the first 5\,yrs
and decelerated from then on. The size estimate {\bf F99} shown is
compatible with this expansion scenario and it would be slightly
different for scenarios where the free expansion would last much
longer than 5\,yr, which is not the case. See text.
\label{fig:fit2points}}
\end{figure}

An estimate of $m$ ($R\propto t^m$) using the size estimate of
\cite{bar85} and our size estimate results in
$m=0.62^{+0.22}_{-0.17}$. This results shows that
\object{SN\,1979C} has indeed dramatically decreased its growth
rate after a likely initial phase of free expansion ($m \approx
1.0$; see Fig.\ \ref{fig:fit2points}). The optical data, although
as scarce as the VLBI data, can be used as a check of the
reliability of our estimate of $m$. In a self-similar expansion
scenario $R\propto t^m$ and, consequently, $v\propto t^{m-1}$. If
we now assume that the supernova did not decelerate for the first
5.17 years, we can take the value of 9\,200 \,km\,s$^{-1}$ as the
expansion velocity just prior to the phase of strongly decelerated
expansion. Combining this value with that of 6\,200$\pm$300
\,km\,s$^{-1}$ at epoch 14.09 \,yr (F99 and R. Fesen, priv.
comm.), gives $m=0.61^{+0.10}_{-0.11}$, in agreement with our
estimate of $m=0.62^{+0.22}_{-0.17}$.

Hence, $m=0.62$ appears to be an adequate value to characterize a
phase of strong deceleration in the expansion of SN\,1979C,
assuming it started about 5.17 years after its explosion. If the
supernova started to decelerate later than 5.17yr, then $m$ should
be smaller, and correspondingly stronger the deceleration of
SN\,1979C. It seems somewhat arbitrary to assume that the
deceleration set in at the same time as the observations of B85.
Indeed, there is nothing apparent in the integrated radio light
curves which could point to a change at that particular epoch. It
is intriguing, though, that the integrated radio light curves of
SN\,1979C (Fig. 1) seem to show an increase in flux at around
t=2200 days, or 6 yr after its explosion. One could speculate that
the flux rise at this epoch is somehow related to a change in the
deceleration parameter, likely due to the shock front of the
supernova entering a high-density region of the circumstellar
medium (CSM).

Our estimate of the angular diameter for SN\,1979C can also be
used to obtain an upper limit for the epoch at which the phase of
strong deceleration started. We find that the latest epoch of free
expansion which is still compatible with both the optical
observations of F99 (epoch 14.09yr) and our VLBI determination at
20.12 yr, and with the fact that a lower limit is set by SN\,1979C
entering a Sedov phase ($m=0.4$), corresponds to $\sim$8 yr.

The integrated radio emission curves have been well modeled for
the first 10 years by  Weiler \et\ (\cite{wei92}) and Montes \et\
(\cite{mon00}) who assumed self-similar expansion and a modulation
in the density of the CSM due to the binary interaction, to
explain the modulation in the flux density curve. Montes \et\
(\cite{mon00}) found that starting around epoch 10\,yr the radio
emission started to grow after a long decline predicted by the
radio emission model. Can our VLBI observations shed any light on
what happened at epoch 10\,yr? The VLBI results by \cite{bar85}
and us require a strong deceleration of the expansion. Such
deceleration is compatible with the optical results by F99 for a
self-similar expansion. Our results rule out any initial free
expansion longer than 8 years. Thus, the onset of decelerated
expansion took place well before the epoch 10\,yr where the radio
emission decline turned onto a slight growth. The sooner the onset
of the decelerated expansion the milder the deceleration needed to
be compatible with the observations. The epoch of that onset is
restricted to the range 4-8 years and the corresponding
deceleration parameter $m$ to the range 0.67-0.44. (Strictly
speaking an onset at 3 years with a corresponding $m$=0.72 is not
excluded, but it appears unlikely since it yields a radio to
optical velocity ratio of 0.68 at epoch 5.17 yr. Such a ratio is
small,  but not yet inconsistent with the models used. In any
case, the deceleration required for an early onset is comparable
to the strongest decelerations reported in other cases by Marcaide
\et\ (\cite{mar97}) for SN\,1993J in M81 and Mc Donald \et\
(\cite{mcd01}) for 43.31+592 in M82). An onset of decelerated
expansion at epoch 6\,yr (corresponding to a deceleration
parameter $m$=0.57) appears most likely and it may be associated
with a small halt in the otherwise decreasing flux density trend,
a halt which may be partly related to encountering a denser CSM.

If the expansion of SN\,1979C has significantly decelerated, as
our VLBI estimate shows, then the mass of the CSM swept up by the
shock front, $M_{\rm sw}$, must be comparable to or larger than
the mass of ejected hydrogen-rich envelope, $M_{\rm env}$. Our
estimated angular diameter translates into a forward shock radius
of $\sim 4.33 \times 10^{17}$ cm. For a presupernova wind of
velocity $v_{\rm w}$=10\,km\,s$^{-1}$, this radius corresponds to
$\sim$13\,700\,yr prior to the explosion of \object{SN\,1979C}.
Weiler et al. (1991) estimated a mass-loss rate of $\sim1.2 \times
10^{-4} M_{\sun}$ yr$^{-1}$ for the progenitor of
\object{SN\,1979C}. If we now assume that the circumstellar medium
up to this distance is a standard one ($\rho_{\rm cs} \propto
r^{-2}$), we then obtain $M_{\rm sw} \sim 1.6 M_{\sun}$. Momentum
conservation therefore implies that $M_{\rm env} \approx 0.59
M_{\rm sw}$, which results in a mass for the ejected hydrogen-rich
envelope of $\sim 0.94 M_{\sun}$.

In the case of the Type II supernova SN1993J, whose progenitor was
a $\sim 15 M_{\sun}$ star in a binary system, mass exchange with
its companion meant that the mass of the hydrogen-rich envelope
ejected at the explosion was $0.2 - 0.4 M_{\sun}$ (Woosley et al.
1994, Houck \& Fransson 1996), much less than the $\approx 3.3
M_{\sun}$ one would expect from a single-star model (H\"oflich  et
al. 1993). Weiler et al. (1986) have suggested that SN1979C was
born in a binary system consisting of a 15-18 $M_{\sun}$ red
supergiant and a 10 $M_{\sun}$ B1 main-sequence star. If we assume
that SN~1979C lost, like SN~1993J, most of its hydrogen-rich
envelope prior to explosion, then the $\sim 0.9 M_{\sun}$
decelerated ejecta suggested from our observations represents at
least all of the remaining hydrogen-rich envelope.

>From VLBI observations of SN\,1979C more than 20 years after its
explosion, we conclude that an initial free expansion of
\object{SN\,1979C} was followed by a strong deceleration which set
on $\sim$6\,yr after the supernova explosion. We estimate the
shocked swept-up mass to be $M_{\rm sw} \sim1.6 M_{\sun}$,
assuming a standard density profile for the CSM. Such a swept-up
mass for \object{SN\,1979C} suggests a mass of the hydrogen-rich
envelope ejected at explosion as low as $M_{\rm env} \sim0.94
M_{\sun}$ or lower. If, as suggested by Weiler et al.
(\cite{wei86}) and Schwarz \& Pringle (\cite{sch96})
\object{SN\,1979C} originated in a binary star (alike
\object{SN\,1993J}) this value of $M_{\rm env}$ suggests that the
companion of the progenitor star stripped off most of the
hydrogen-rich envelope mass of the presupernova star prior to its
explosion. New observations at $\lambda$18\,cm around 2005 should
be able to test the above conclusion and would also be essential
to image the true structure of \object{SN\,1979C}. On the other
hand, a continued rise in the level of flux density at
$\lambda$6\,cm and improved VLBI instrumentation and techniques
like phase-referencing may allow an earlier determination of the
intrinsic structure of \object{SN\,1979C} with high resolution and
a test of the binary star scenario which the present results tend
to favor.

\begin{acknowledgements}
We are grateful to A.\ Kraus for the help provided during the
observations and the flux density measurements on
\object{PKS\,B1157+156}. This work has been partially supported by
the Spanish DGICYT Grants No.\ PB96-0782 and PB97-1164.  NRAO is
operated under license by Associated Universities, Inc., under
cooperative agreement with NSF. We acknowledge support from the
European Commission's TMR-LSF programme, Contract No.
ERBFMGEST950012. JMM is grateful to the Max-Planck Institut f\"ur
Radioastronomie for the support received for this research during
his visit. KWW wishes to thank the Office of Naval Research for
the 6.1 funding supporting this work.

\end{acknowledgements}

\end{document}